\documentclass[12pt]{article}
\pdfoutput=1                          
\usepackage[utf8x]{inputenc}
\usepackage[english]{babel}

\usepackage{ucs}
\usepackage{amsmath}
\usepackage{amsfonts}
\usepackage{amssymb}
\usepackage{eucal}               
\usepackage{mathrsfs}          
\usepackage[sans]{dsfont}     
\usepackage[Symbol]{upgreek}  

\usepackage{array}             
\usepackage{booktabs}       
\usepackage{colortbl}          
\usepackage{multirow}
\usepackage{longtable}

\usepackage{cite}                

\usepackage{fancyhdr}

\usepackage[normalem]{ulem}             

\usepackage{cancel}
\usepackage{slashed}
\usepackage{simplewick}     
\usepackage{feynmf}           

\usepackage{bm}
\usepackage{mathtools}
\DeclareMathAlphabet{\mathpzc}{OT1}{pzc}{m}{it}
\DeclareMathAlphabet{\matheul}{U}{eus}{m}{n}

\usepackage
{graphicx}

\usepackage{color}
\definecolor{grigio}{cmyk}{0,0,0,0.1}
\definecolor{rosa}{cmyk}{0,0.1,0.1,0.02}
\definecolor{rosino}{cmyk}{0,0.05,0.05,0.02}
\definecolor{rosas}{cmyk}{0,0.3,0.25,0.05}
\definecolor{celeste}{cmyk}{0.1,0,0,0.02}
\definecolor{giallino}{cmyk}{0,0,0.1,0.02}
\definecolor{rosso}{cmyk}{0,1,1,0.4}
\definecolor{rossos}{cmyk}{0,1,1,0.55}
\definecolor{rossoc}{cmyk}{0,1,1,0.2}
\definecolor{blu}{cmyk}{1,1,0,0.3}
\definecolor{blus}{cmyk}{1,1,0,0.5}
\definecolor{bluc}{cmyk}{1,1,0,0.1}
\definecolor{blucc}{cmyk}{0.7,0.5,0,0}
\definecolor{viola}{cmyk}{0,1,0,0.6}
\definecolor{viola2}{cmyk}{0,1,0.2,0.6}
\definecolor{verde}{cmyk}{0.92,0,0.59,0.25}
\definecolor{verdec}{cmyk}{0.92,0,0.59,0.15}
\definecolor{verdes}{cmyk}{0.92,0,0.59,0.4}
\definecolor{verdino}{cmyk}{0.12,0,0.09,0.02}
\definecolor{giallo}{cmyk}{0,0,1,0}
\definecolor{gialloverde}{cmyk}{0.44,0,0.74,0}
\definecolor{Titolo}{rgb}{0.752941176,0.576470588,0.992156863}
\definecolor{altro}{rgb}{0.094117647,0.650980392,0.643137255}
\definecolor{Peanuts}{rgb}{0.2, 0.4, 0.6}
\definecolor{Pean1}{rgb}{0.6, 0.8, 0.4}
\definecolor{BHO}{rgb}{0.2, 0.8, 1}
\definecolor{Daria}{rgb}{0, 0.9412, 0}
\definecolor{UniPi}{rgb}{0.2549, 0.4627, 0.6275}
\definecolor{UniPidue}{rgb}{0.3216, 0.5804, 0.7882}
\definecolor{rossoCP3}{cmyk}{0,.88,.77,.40}
\definecolor{verdeCP3}{rgb}{0.09765625, 0.57421875, 0.1015625}
\definecolor{bluCP3}{rgb}{0, 0.23, 0.67}

\newcommand{\spazio}{\bigskip}

\newcommand{\virg}[1]{`#1'}

\newcommand{\eg}{e.g.~}
\newcommand{\ie}{i.e.~}

\newcommand{\Eq}[1]{Eq.~\eqref{#1}}
\newcommand{\Fig}[1]{Fig.~\ref{#1}}
\newcommand{\Sec}[1]{Sec.~\ref{#1}}

\newcommand{\Lag}{\mathscr{L}}	

\newcommand{\beq}{\begin{equation}}
\newcommand{\eeq}{\end{equation}}
\newcommand{\ud}{\text{d}}

\newcommand{\mDM}{m}
\newcommand{\ER}{E_\text{R}}
\newcommand{\Ed}{E'}

\newcommand{\vmin}{v_\text{min}}

\newcommand{\bfv}{\mathbf{v}}
\newcommand{\eH}{\matheul{H}}
\newcommand{\eR}{\matheul{R}}

\long\def\symbolfootnote[#1]#2{\begingroup\def\thefootnote{\fnsymbol{footnote}}\footnote[#1]{#2}\endgroup}

\ifx\pdfoutput\undefined
\usepackage[dvips,bookmarks]{hyperref}    
\else
\usepackage{hyperref}    
\fi

\def\mhref#1{\href{mailto:#1}{#1}}        

\newlength{\biblength}
\makeatletter
\renewcommand\@biblabel[1]{\parbox[c]{\biblength}{\hfill[#1]}}		
\makeatother

\begin{document}

\begin{titlepage}

\begin{center}
{\Large{\bf Generalized Halo Independent Comparison\\[2mm] of Direct Dark Matter Detection Data}}
\end{center}

\par \vskip .2in \noindent

\begin{center}
{\sc
Eugenio Del Nobile$\, {^{1}}$\,\symbolfootnote[2]{\mhref{delnobile@physics.ucla.edu}},
Graciela Gelmini$\, {^{1}}$\,\symbolfootnote[1]{\mhref{gelmini@physics.ucla.edu}},
\\
Paolo Gondolo$\, {^2}$\,\symbolfootnote[1]{\mhref{paolo.gondolo@utah.edu}},
Ji-Haeng Huh$\, {^1}$\,\symbolfootnote[3]{\mhref{jhhuh@physics.ucla.edu}}}
\end{center}

\begin{center}
\par \vskip .1in \noindent
{\it ${^1} \, $Department of Physics and Astronomy, UCLA,\\ 
475 Portola Plaza, Los Angeles, CA 90095, USA}
\par \vskip .1in \noindent
{\it ${^2} \, $Department of Physics and Astronomy, University of Utah,\\
115 S 1400 E Suite 201, Salt Lake City, UT 84112, USA}
\par \vskip .1in \noindent
\end{center}

\begin{center}
{\large Abstract}
\end{center}

\begin{quote}
We extend the halo-independent method to compare direct dark matter detection data, so far used only for spin-independent WIMP-nucleon interactions, to any type of interaction. As an example we apply the method to magnetic moment interactions.
\end{quote}

\end{titlepage}

\newpage


\section{Introduction}
Determining what the dark matter (DM), the most abundant form of matter in the Universe, consists of is one of the most fundamental open questions in physics and cosmology. Weakly interacting massive particles (WIMPs) are among the most experimentally sought after candidates. Four direct detection experiments, DAMA~\cite{Bernabei:2010mq}, CoGeNT~\cite{Aalseth:2010vx, Aalseth:2011wp}, CRESST-II~\cite{Angloher:2011uu} and CDMS-II-Si \cite{Agnese:2013rvf} have reported potential signals of WIMP DM, while all other direct detection searches have produced only upper bounds on interaction rates and annual modulation of the signal \cite{Angle:2011th, Aprile:2011hi, Aprile:2012nq, Felizardo:2011uw, Ahmed:2010wy, 
Ahmed:2012vq, Agnese:2013cvt}.

An interesting way of comparing all these data, which circumvents the uncertainties in our knowledge of the local characteristics of the dark halo of our galaxy, is the \virg{halo-independent} comparison method \cite{Fox:2010bz, Frandsen:2011gi, Gondolo:2012rs, Frandsen:2013cna, DelNobile:2013cta, HerreroGarcia:2011aa, HerreroGarcia:2012fu, Bozorgnia:2013hsa}. The main idea of this method is that the interaction rate at one particular recoil energy $\ER$ depends for any experiment on one and the same  function $\eta(\vmin)$ of the minimum speed $\vmin$ required for the incoming DM particle to cause a nuclear recoil with energy $\ER$. The function $\eta(\vmin)$  depends only on the local characteristics of the dark  halo of our galaxy. Thus, all rate measurements and bounds can be translated into measurements and bounds on the unique function $\eta(\vmin)$.

So far, this method was applied to the standard spin-independent (SI) WIMP-nucleus interaction  only, although it could easily be applied to the standard spin-dependent (SD) interaction as well.  For both SI and SD interactions, the differential scattering cross section has a $1/v^2$ dependence on the speed $v$ of the DM particle. However, there are many other kinds of interactions with more general dependence  on the DM particle velocity and on the nuclear recoil energy. Examples of these are: DM interacting through effective operators~\cite{Bagnasco:1993st, Kurylov:2003ra, Dobrescu:2006au, Fan:2010gt, Fitzpatrick:2012ix, Beltran:2008xg, Chang:2009yt, Zheng:2010js, Yu:2011by, Cheung:2012gi, MarchRussell:2012hi, Ding:2012sm},  WIMPs with electromagnetic couplings~\cite{Pospelov:2000bq, Fitzpatrick:2010br, An:2010kc, McDermott:2010pa}, in particular via a magnetic dipole moment~\cite{Sigurdson:2004zp, Barger:2010gv, Chang:2010en, Cho:2010br, Heo:2009vt, Gardner:2008yn, Masso:2009mu, Banks:2010eh, Fortin:2011hv, Kumar:2011iy, Barger:2012pf, DelNobile:2012tx, Cline:2012is, Weiner:2012cb, Tulin:2012uq, Cline:2012bz}, ``Resonant DM"~\cite{Pospelov:2008qx, Bai:2009cd}, ``Form factor DM"~\cite{Feldstein:2009tr}, ``Anapole DM"~\cite{Ho:2012bg}.  While some of these interactions  can be treated in a halo-independent fashion with trivial modifications of the method used so far for the SI interactions, this method cannot be applied to all of them, such as the magnetic dipole and anapole moment interactions, or resonant DM.

  What forbids a trivial extension of the SI method in the case of magnetic dipole and anapole moment interactions is that the cross sections contain two different terms with different dependences on the DM particle speed $v$. When these terms are integrated over the velocity distribution to find the rate, instead of a unique function $\eta(\vmin)$, each term has its own function of $\vmin$ multiplied by its own detector dependent coefficient. It is thus impossible  to translate a rate measurement or bound into only one of the two $\vmin$ functions contributing to the rate. Similarly, what forbids a trivial extension of the SI method to the case of ``Resonant DM'' is that the cross section for ``Resonant DM" has a Breit-Wigner energy dependence with a shape that depends on the target nucleus. Thus each target has its own function of $\vmin$, and again it seems impossible to find one and the same common function analogous to $\eta(\vmin)$ so that all rate measurements and bounds can be mapped onto it.

The aim of this paper is to extend the halo-independent analysis to all interactions circumventing the complications just mentioned. We find for any kind of interaction how to map all the rate measurements and bounds obtained with different experiments into a unique function of $\vmin$ that depends on the local characteristics of the dark halo of our galaxy only.

In \Sec{haloindep-SI} we fix our notation and recall the formalism for the halo-independent analysis as used so far for SI interactions. In \Sec{haloindep} we present our generalized halo independent method, applicable to any type of interaction. We  concentrate on elastic collisions, but present the formalism for inelastic collisions in Appendix A. Then in \Sec{measurements} we explain how we proceed to compare data in a halo independent manner, and in Secs.~5 and 6  we apply the method to Magnetic Dipole Moment DM (MDM). In Sec.~7 we present our concluding remarks.

\section{Halo-independent method - SI interactions}\label{haloindep-SI}

The DM-nucleus differential scattering rate in counts/kg/day/keV for nuclear recoil energy $\ER$ and  target nuclide $T$, is
\beq
\label{dRdER}
\frac{\ud R_T}{\ud E_\text{R}} = \frac{\rho}{\mDM} \frac{C_T}{m_T} \int_{v \geqslant v_\text{min}(\ER)} \hspace{-24pt} \ud^3 v \, f(\bfv, t) \, v \, \frac{\ud \sigma_T}{\ud \ER}(\ER, \bfv) .
\eeq
 Here $m$ is the DM particle mass, $m_T$ is the target nuclide mass, and $C_T$ is mass fraction of nuclide $T$ in the detector. The dependence of the rate on the local characteristics of the dark halo is contained in the local DM density $\rho$ and the DM velocity distribution in the Earth's frame $f(\bfv, t)$, which is modulated in time due to the Earth's rotation around the Sun. The distribution $f(\bfv, t)$ is normalized to $\int \ud^3 v \, f(\bfv, t) = 1$. The minimum speed required for the incoming DM particle to cause a nuclear recoil with energy $\ER$ is $\vmin$. For an elastic collision (see Appendix A for inelastic collisions),
\beq\label{vmin}
\vmin = \sqrt{\frac{m_T \ER}{2 \mu_T^2}} ,
\eeq
 where $\mu_T=m~m_T/(m+m_T)$ is the WIMP-nucleus reduced mass.

To properly reproduce the recoil rate measured by experiments, we need to take into account the characteristics of the detector. Most experiments do not measure the recoil energy directly but  rather a detected energy $\Ed$, often quoted in keVee (keV electron-equivalent) or in photoelectrons. The uncertainties and fluctuations in the detected energy corresponding to a particular recoil energy are expressed  in a (target nuclide and detector dependent) resolution function $G_T(\ER, \Ed)$ that gives the probability that a recoil energy $\ER$ is measured as $\Ed$.  The resolution function is often but not always (the XENON experiments are a notable exception) approximated by a Gaussian distribution. It incorporates the mean value $\langle \Ed \rangle = Q_T \ER$, which depends on  the energy dependent quenching factor $Q_T(\ER)$, and the energy resolution $\sigma_{\ER}(\Ed)$.  Moreover, experiments have a counting efficiency or cut acceptance $\epsilon(\Ed)$, which also affects the measured rate.
Thus the nuclear recoil rate in \Eq{dRdER} must be convolved with the function $\epsilon(\Ed) G_T(\ER, \Ed)$. The resulting differential rate as a function of the detected energy $\Ed$ is
\beq\label{dRdEd}
\frac{\ud R}{\ud \Ed} = \epsilon(\Ed) \sum_T \int_0^\infty \ud \ER \, G_T(\ER, \Ed) \frac{\ud R_T}{\ud \ER} .
\eeq
The rate within a detected energy interval $[ \Ed_1, \Ed_2]$  follows as
\begin{multline}
\label{R}
R_{[\Ed_1, \Ed_2]}(t) =  \int_{\Ed_1}^{\Ed_2} \ud\Ed \, \frac{\ud R}{\ud \Ed} 
\\
= \frac{\rho}{\mDM} \sum_T \frac{C_T}{m_T} \int_0^\infty \ud \ER \, \int_{v \geqslant v_\text{min}(\ER)} \hspace{-18pt} \ud^3 v \, f(\bfv, t) \, v \, \frac{\ud \sigma_T}{\ud \ER}(\ER, \bfv)
\\
\times
 \int_{\Ed_1}^{\Ed_2} \ud\Ed \, \epsilon(\Ed) G_T(\ER, \Ed).
\end{multline}

The differential cross section for the usual SI interaction is
\beq
\frac{\ud \sigma_T}{\ud \ER} = \sigma_T^{\rm SI}(\ER) \frac{m_T}{2 \mu_T^2 v^2} ,
\eeq
with
\beq
\sigma_T^{\rm SI}(\ER) = \sigma_p \frac{\mu_T^2}{\mu_p^2} | Z_T + (A_T - Z_T) f_n / f_p |^2 F_{{\rm SI}, T}^2(\ER) .
\eeq
 Here $Z_T$ and $A_T$ are respectively the atomic and mass number of the target nuclide $T$, $F_{{\rm SI}, T}(\ER)$ is the nuclear spin-independent form factor, $f_n$ and $f_p$ are the effective DM couplings to neutron and proton, and $\mu_p$ is the DM-proton reduced mass. Using this expression for the differential cross section, and changing integration variable from $\ER$ to $\vmin$ through \Eq{vmin}, we can rewrite \Eq{R} as
\begin{align}
\label{R1}
R^{\rm SI}_{[\Ed_1, \Ed_2]}(t) & =  \int_0^\infty \ud \vmin \, \tilde{\eta}(\vmin, t) \, \eR^{\rm SI}_{[\Ed_1, \Ed_2]}(\vmin) ,
\end{align}
where the velocity integral $\tilde{\eta}$ is 
\beq
\label{eta0}
\tilde{\eta}(\vmin, t) \equiv \frac{\rho \sigma_p}{\mDM} \int_{v \geqslant \vmin} \ud^3 v \, \frac{f(\bfv, t)}{v} \equiv \int_{\vmin}^\infty \ud^3 v \, \frac{\tilde{f}(\bfv, t)}{v} ,
\eeq
and we defined the response function $\eR^{\rm SI}_{[\Ed_1, \Ed_2]}(\vmin)$ for WIMPS with SI interactions as
\begin{multline}
\eR^{\rm SI}_{[\Ed_1, \Ed_2]}(\vmin) \equiv
2 \vmin \sum_T \frac{C_T}{m_T} \frac{\sigma_T^{\rm SI}(\ER(\vmin))}{\sigma_p}
\\
\times
 \int_{\Ed_1}^{\Ed_2} \ud\Ed \, \epsilon(\Ed) G_T(\ER(\vmin), \Ed).
\end{multline}
Introducing the speed distribution
\begin{align}
\widetilde{F}(v, t) \equiv v^2 \int \ud\Omega_v \, \tilde{f}(\bfv, t) ,
\end{align}
we can rewrite the $\tilde{\eta}$ function as
\beq\label{etaF}
\tilde{\eta}(\vmin, t) = \int_{\vmin}^\infty \ud v \, \frac{\widetilde{F}(v, t)}{v} .
\eeq

Due to the revolution of the Earth around the Sun,  the velocity integral  $\tilde{\eta}(\vmin,t)$ has an annual modulation generally well approximated by the first terms of a harmonic series,
\beq\label{etat}
\tilde{\eta}(\vmin, t) \simeq \tilde{\eta}^0(\vmin) + \tilde{\eta}^1(\vmin) \cos\!\left[ \omega (t - t_0) \right],
\eeq
where $t_0$ is the time of the maximum of the signal and $\omega = 2 \pi/$yr. The unmodulated and modulated components $\tilde{\eta}^0$ and $\tilde{\eta}^1$ enter respectively in the definition of unmodulated and modulated parts of the rate,
\beq
R_{[\Ed_1, \Ed_2]}(t) = R^0_{[\Ed_1, \Ed_2]} + R^1_{[\Ed_1, \Ed_2]} \cos\!\left[ \omega (t - t_0) \right] .
\eeq
Once the WIMP mass and interactions are fixed, the functions $\tilde{\eta}^0(\vmin)$ and $\tilde{\eta}^1(\vmin)$ are detector-independent quantities that must be common to all non-directional direct dark matter experiments. Thus we can map the rates measurements and bounds of different experiments  into measurements of and bounds on $\tilde{\eta}^0(\vmin)$ and $\tilde{\eta}^1(\vmin)$ as functions of $\vmin$.

Averages of the $\tilde{\eta}^i$ functions weighted by the response function $\eR^{\rm SI}_{[\Ed_1, \Ed_2]}(\vmin)$ were compared in Refs.~\cite{Gondolo:2012rs} and \cite{DelNobile:2013cta} with upper limits on $\tilde{\eta}^i$. The weighted averages practically coincide with the values assigned to the $\tilde{\eta}^i$ functions in Refs.~\cite{Fox:2010bz},  \cite{Frandsen:2011gi} and \cite{Frandsen:2013cna} when, as assumed in  those references, the energy interval is small enough that the differential rate, form factor and efficiency can be taken to be constant  within the interval.  

 For experiments with putative DM signals, a rate $\hat{R}^{\, i}_{[\Ed_1, \Ed_2]}$ measured by an experiment in an energy interval $[\Ed_1, \Ed_2]$, translates into the average of $\tilde{\eta}^{\, i}(\vmin)$ in the corresponding $\vmin$ interval $[{\vmin}_{,1}, {\vmin}_{,2}]$ in which the response function
 $\eR^{\rm SI}_{[\Ed_1, \Ed_2]}(\vmin)$ is sufficiently different from zero,
\beq
\label{avereta}
\overline{\tilde{\eta}^{\, i}_{[\Ed_1, \Ed_2]}} \equiv \frac{\hat{R}^{\, i}_{[\Ed_1, \Ed_2]}}
{\int \ud\vmin \, \eR^{\rm SI}_{[\Ed_1, \Ed_2]}(\vmin)} ,
\eeq
with $i = 0, 1$ for the unmodulated and modulated component, respectively.   The interval $[{\vmin}_{,1}, {\vmin}_{,2}]$ determines  the width of the horizontal ``error bar''  in the $(\vmin, \tilde{\eta})$ plane.
 In practice,  following Ref.~\cite{Frandsen:2011gi}, for simplicity  ${\vmin}_{,1}$ and ${\vmin}_{,2}$ were so far approximated by $v_{\rm min,1} = v_{\rm min}(E'_1-\sigma_{\ER}(E'_1))$ and $v_{\rm min,2} = v_{\rm min}(E'_2+\sigma_{\ER}(E'_2))$.
The vertical ``error bar,'' unless otherwise indicated, showed the $68\%$ confidence interval with Poissonian statistics.

To determine the upper bounds on the unmodulated part of $\tilde{\eta}$ set by  experimental upper bounds on the unmodulated part of the rate, the procedure first outlined in Refs.~\cite{Fox:2010bz, Frandsen:2011gi} was used.  This limit exploits the fact that by definition  $\tilde{\eta}^0$ is a non-increasing function of $\vmin$, thus  the smallest possible $\tilde{\eta}^0(\vmin)$ function passing by a fixed point $(v_0, \tilde{\eta}_0)$ in the $(\vmin, \tilde{\eta})$ plane, is the downward step-function $\tilde{\eta}_0 \, \theta(v_0 - \vmin)$.  In other words, among the functions passing by the point $(v_0, \tilde{\eta}_0)$, the downward step is the function yielding the minimum predicted number of events. Imposing this functional form in \Eq{R1}
\beq
R_{[\Ed_1, \Ed_2]} = \tilde{\eta}_0 \int_0^{v_0} \ud \vmin \, \eR^{\rm SI}_{[\Ed_1, \Ed_2]}(\vmin) .
\eeq
The upper bound $R^{\rm lim}_{[\Ed_1, \Ed_2]}$ on the unmodulated rate in an interval $[\Ed_1, \Ed_2]$ (usually at the $90\%$ confidence level) is translated into an upper bound $\tilde{\eta}^{\rm lim}(\vmin)$ on $\tilde{\eta}^0$ at $v_0$ by
\beq
\tilde{\eta}^{\rm lim}(v_0) = \frac{R^{\rm lim}_{[\Ed_1, \Ed_2]}}{\int_0^{v_0} \ud \vmin \, \eR^{\rm SI}_{[\Ed_1, \Ed_2]}(\vmin)} .
\eeq
The upper bound so obtained is conservative in the sense that there are excluded functions $\tilde{\eta}^0(\vmin)$ that nowhere  exceed the limit. In other words, all functions $\tilde{\eta}^0(\vmin)$ for which $\tilde{\eta}^0(\vmin) > \tilde{\eta}^{\rm lim}(\vmin)$ at some $\vmin$ are excluded, but there are other excluded functions for which $\tilde{\eta}^0(\vmin) \le \tilde{\eta}^{\rm lim}(\vmin)$ at all $\vmin$ \cite{Frandsen:2011gi}.

 The procedure just described does not assume any particular property of the dark halo. By making some assumptions, more stringent limits on the modulated part $\tilde{\eta}^1$ can be derived  from the limits on the unmodulated part of the rate (see Refs.~\cite{HerreroGarcia:2011aa, HerreroGarcia:2012fu, Bozorgnia:2013hsa}), but we choose to proceed without making any assumption on the dark halo.

The procedure outlined in this section to compare data from different experiments in a halo independent way  can only be applied when the differential cross section can be factorized into a velocity dependent term, independent of the detector (\eg it must be independent of $m_T$), times a velocity independent term containing all the detector dependency. In the case of a more general form of the differential cross section, we can instead proceed as described in the following section.

\section{Generalized halo independent method}\label{haloindep}

 Here we present a way of defining the response function $\eR_{[\Ed_1, \Ed_2]}(\vmin)$ in \Eq{R1} that is valid for any type of interaction. Changing the order of the $\bfv$ and $\ER$ integrations in \Eq{R}, we have
\begin{multline}\label{R2}
R_{[\Ed_1, \Ed_2]}(t) =
\frac{\rho \sigma_{\rm ref}}{\mDM} \int_0^\infty \ud^3 v \, \frac{f(\bfv, t)}{v}
\sum_T \frac{C_T}{m_T} \int_0^{\ER^{\rm max}(v)} \ud \ER \, \frac{v^2}{\sigma_{\rm ref}} \frac{\ud \sigma_T}{\ud \ER}(\ER, \bfv)
\\
\times
 \int_{\Ed_1}^{\Ed_2} \ud\Ed \, \epsilon(\Ed) G_T(\ER, \Ed) .
\end{multline}
Here $\ER^{\rm max}(v) \equiv 2 \mu_T^2 v^2 / m_T$ is the maximum recoil energy a WIMP of speed $v$ can impart  in an elastic collision to a target nucleus $T$ initially at rest. To make contact with the SI interaction method of the previous section, we have multiplied and divided by the factor $\sigma_{\rm ref} / v^2$, where $\sigma_{\rm ref}$ is a target-independent reference cross section (\ie a constant with the dimensions of a cross section) that coincides with $\sigma_p$  for SI interactions. In compact form, \Eq{R2} reads
\beq
R_{[\Ed_1, \Ed_2]}(t) =  \int_0^\infty \ud^3 v \, \frac{\tilde{f}(\bfv, t)}{v} \, \eH_{[\Ed_1, \Ed_2]}(\bfv) ,
\eeq
where in analogy with \Eq{eta0} we defined
\beq
\label{ftilde}
\tilde{f}(\bfv, t) \equiv \frac{\rho \sigma_{\rm ref}}{\mDM} \, f(\bfv, t) ,
\eeq
and we defined the ``integrated response function" (the name stemming from \Eq{eq:RT})
\begin{multline}
\label{eq:HT}
\eH_{[\Ed_1, \Ed_2]}(\bfv) \equiv
\sum_T \frac{C_T}{m_T} \int_0^{\ER^{\rm max}(v)} \ud \ER \, \frac{v^2}{\sigma_{\rm ref}} \frac{\ud \sigma_T}{\ud \ER}(\ER, \bfv)
\\
\times
 \int_{\Ed_1}^{\Ed_2} \ud\Ed \, \epsilon(\Ed) G_T(\ER, \Ed) \  .
\end{multline}

It will prove useful later to rewrite $\eH_{[\Ed_1, \Ed_2]}$ by changing integration variable from $\ER$ to $\vmin$ through \Eq{vmin}, which yields
\begin{multline}
\label{eq:HT2}
\eH_{[\Ed_1, \Ed_2]}(\bfv) =
\sum_T \frac{C_T}{m_T} \frac{4 \mu_T^2}{m_T} \int_0^v \ud \vmin \, \vmin \frac{v^2}{\sigma_{\rm ref}} \frac{\ud \sigma_T}{\ud \ER}(\ER(\vmin), \bfv)
\\
\times
 \int_{\Ed_1}^{\Ed_2} \ud\Ed \, \epsilon(\Ed) G_T(\ER(\vmin), \Ed) .
\end{multline}

 For simplicity, we only consider differential cross sections, and thus integrated response functions, that depend only on the speed $v=|\bfv|$, and not on the whole velocity vector. This is true if the DM flux and the target nuclei are unpolarized and the detection efficiency is isotropic throughout the detector, which is the most common case. With this restriction,
\begin{align}
R_{[\Ed_1, \Ed_2]}(t) & =  \int_0^\infty \ud v \, \frac{\widetilde{F}(v, t)}{v} \, \eH_{[\Ed_1, \Ed_2]}(v) .
\label{eq:REEbis}
\end{align}
We now define the function $\tilde{\eta}(v, t)$ by
\begin{align}
\label{eta-derivative}
\frac{\widetilde{F}(v, t)}{v}  = - \frac{ \partial \tilde{\eta}(v, t) }{\partial v} ,
\end{align}
with $\tilde{\eta}(v, t)$ going to zero in the limit of $v$ going to infinity. This yields the usual definition of $\tilde{\eta}$ (see \Eq{etaF})
\begin{align}
\label{tildeeta}
\tilde{\eta}(v, t) = \int_v^\infty \ud v' \, \frac{\widetilde{F}(v', t)}{v'} .
\end{align}
Using Eq.~(\ref{eta-derivative}) in Eq.~(\ref{eq:REEbis})
the energy integrated rate becomes
\begin{align}
\label{R3}
R_{[\Ed_1, \Ed_2]}(t) & = - \int_0^\infty \ud v \, \frac{ \partial \tilde{\eta}(v, t) }{\partial v}  \, \eH_{[\Ed_1, \Ed_2]}(v) .
\end{align}
Integration by parts of Eq.~(\ref{R3}) leads to an equation  formally identical to \Eq{R1} but which is now valid for any interaction,
\begin{align}
\label{R4}
R_{[\Ed_1, \Ed_2]}(t) & =  \int_0^\infty \ud v \, \tilde{\eta}(v, t) \,  \eR_{[\Ed_1, \Ed_2]}(v) .
\end{align}
The response function is now defined as the derivative of the ``integrated response function" $\eH_{[\Ed_1, \Ed_2]}(v)$
\begin{align}
\eR_{[\Ed_1, \Ed_2]}(v) \equiv \frac{\partial \eH_{[\Ed_1, \Ed_2]}(v)}{\partial v} .
\label{eq:RT}
\end{align}
Notice that the boundary term in the integration by parts of \Eq{R3} is zero because the definition of $\eH_{[\Ed_1, \Ed_2]}(\bfv)$ in Eq.~(\ref{eq:HT}) imposes that $\eH_{[\Ed_1, \Ed_2]}(0) = 0$ (since $\ER^{\rm max}(0) = 0$).

\spazio

In the rest of the paper we will only use the equations presented up to this point. However, it is interesting to notice that other expressions for the rate are possible. In fact, one can continue the integration by parts procedure of Eq.~(\ref{R3}) to get a generalized version of \Eq{R4}. Defining iteratively
\begin{align}
\label{etak}
\tilde{\eta}_{(k)}(v, t) \equiv k  \int_v^\infty \ud v' \, \tilde{\eta}_{(k-1)}(v', t) ,
\end{align}
for $k$ a positive integer, with $\tilde{\eta}_{(0)}(v, t) \equiv \tilde{\eta}(v, t)$, one can repeatedly integrate \Eq{R4} by parts to get
\begin{align}
\label{R5}
R_{[\Ed_1, \Ed_2]}(t) = \int_0^\infty \ud v \, \tilde{\eta}_{(k)}(v, t) \, \eR^{(k)}_{[\Ed_1, \Ed_2]}(v) ,
\end{align}
where we also defined the response function of the $k$-th order 
\begin{align}
\eR^{(k)}_{[\Ed_1, \Ed_2]}(v) \equiv \frac{1}{k!} \,  \frac{\partial^{k} \eR_{[\Ed_1, \Ed_2]}(v)}{\partial v^{k}} .
\end{align}
A derivation of this result is given in Appendix B. All boundary terms of the successive integrations by parts vanish because we have assumed that the response function and all of its derivatives vanish at $v=0$, a reasonable assumption since $v=0$ is below the threshold of any experiment. Going back to the 3-dimensional DM velocity distribution $\tilde{f}(\bfv, t)$, it is easy to prove (see Appendix B) that $\tilde{\eta}_{(k)}(\vmin)$ is the so-called ``$k$-th partial moment" of the function $\tilde{f}(\bfv)/v$, defined as
\begin{align}
\label{partialmoments}
\tilde{\eta}_{(k)}(\vmin, t) = \int_{v \geqslant \vmin} (v - \vmin)^k \, \frac{\tilde{f}(\bfv, t)}{v} \, \ud^3 v .
\end{align}

\section{Measurements of and bounds on $\tilde{\eta}$}
\label{measurements}

For the time being, similarly to what we did earlier for SI interactions, we want again to compare average values of the $\tilde{\eta}^i$ functions with upper limits. However, 
for a differential cross section with a general dependence on the DM velocity, it might not be possible to simply use \Eq{avereta} with $\eR^{\rm SI}_{[\Ed_1, \Ed_2]}$ replaced by $\eR_{[\Ed_1, \Ed_2]}$ to assign a weighted  average of $\tilde{\eta}^0$ or $\tilde{\eta}^1$ to a finite $\vmin$ range. This may happen because the width of the response function $\eR_{[\Ed_1, \Ed_2]}(v)$ in \Eq{eq:RT} at large $\vmin$  is dictated by the high speed behavior of the differential cross section, and it might even be infinite. For example, if $v^2 \, (\ud\sigma_T / \ud\ER)$ goes as $v^n$, with $n$ a positive integer,  for large $v$,
then $\eH_{[\Ed_1, \Ed_2]}(v)$ also goes as $v^n$ and  $\eR_{[\Ed_1, \Ed_2]}(v)$ goes as $v^{n-1}$ for large $v$. Thus, if $n \geqslant 1$, the response function $\eR_{[\Ed_1, \Ed_2]}(v)$ does not vanish for large $v$.  This implies that the denominator in \Eq{avereta} diverges. 

However, we can regularize the behavior of the response function at large $v$ by using for example the function $v^r \tilde{\eta}(v)$ with integer $r \geqslant n$, instead of just $\tilde{\eta}(v)$.  Since this new function is common to all experiments, we can use it to compare the data in $\vmin$ space.\footnote{While any other function that goes to zero fast enough would be equally good to regularize $\eR_{[\Ed_1, \Ed_2]}(v)$, like for instance an exponentially decreasing function, we have chosen the power law $v^{-r}$ because it does not require the introduction of an arbitrary $v$ scale in the problem.} In fact, by multiplying and dividing the integrand in \Eq{R4} by $v^r$, we can define the average of the function $v^r \tilde{\eta}(v)$ with weights $v^{-r} \mathcal{R}_{[E'_1,E'_2]}(v)$,
\beq
\overline{v^r \tilde{\eta}^{\, i}_{[\Ed_1, \Ed_2]}}=  \frac{ \int_0^\infty \ud v \, v^r \, \tilde{\eta}(v, t) \,  v^{-r} \eR_{[\Ed_1, \Ed_2]}(v)}{\int_0^\infty \ud v \, v^{-r} \,  \eR_{[\Ed_1, \Ed_2]}(v)}.
\eeq
With this definition,
\beq
\label{averetavr}
\overline{v^r \tilde{\eta}^{\, i}_{[\Ed_1, \Ed_2]}} = \frac{\hat{R}^{\, i}_{[\Ed_1, \Ed_2]}}{\int^\infty_0 \ud v \, v^{-r} \, \eR_{[\Ed_1, \Ed_2]}(v)} .
\eeq
Notice that exploiting the definition of $\eR_{[\Ed_1, \Ed_2]}$ in \Eq{eq:RT}, we can write this relation in terms of $\eH_{[\Ed_1, \Ed_2]}$ instead of $\eR_{[\Ed_1, \Ed_2]}$ as
\beq
\label{averetavr2}
\overline{v^r \tilde{\eta}^{\, i}_{[\Ed_1, \Ed_2]}} = \frac{\hat{R}^{\, i}_{[\Ed_1, \Ed_2]}}{r \int^\infty_0 \ud v \, v^{-r-1} \, \eH_{[\Ed_1, \Ed_2]}(v)} ,
\eeq
where in the integration by parts the  finite term $\left[ v^{-r} \, \eH_{[\Ed_1, \Ed_2]}(v) \right]^\infty_0$ vanishes since by assumption $r$ has been appropriately chosen to regularize the integral of $v^{-r} \eR_{[\Ed_1, \Ed_2]}(v)$, \ie $v^{-r} \, \eH_{[\Ed_1, \Ed_2]}(v)\to 0$ as $v \to \infty$.

Eqs.~\eqref{averetavr} or \eqref{averetavr2} allow to translate rate measurements in a detected energy interval $[\Ed_1, \Ed_2]$ into averaged values of $v^r \tilde{\eta}(v)$ in a finite $\vmin$ interval $[{\vmin}_{,1}, {\vmin}_{,2}]$. This is now the interval outside which the integral of  the new response function $v^{-r} \, \eR_{[\Ed_1, \Ed_2]}(v)$ (and not of $\eR_{[\Ed_1, \Ed_2]}(v)$) is negligible. We choose to use $90\%$ central quantile intervals, \ie  we determine ${\vmin}_{,1}$ and ${\vmin}_{,2}$ such that the area under the function  $v^{-r} \, \eR_{[\Ed_1, \Ed_2]}(v)$ to the left of ${\vmin}_{,1}$ is $5\%$ of the total area, and the area to the right of ${\vmin}_{,2}$ is also $5\%$ of the total area. In practice, the larger the value of $r$, the smaller is the width of the $[{\vmin}_{,1}, {\vmin}_{,2}]$  interval, designated by  the horizontal ``error bar'' of the crosses in the $(\vmin, \tilde{\eta})$ plane. However, $r$  cannot be chosen arbitrarily large, because  large values of $r$ give a large weight to the low velocity tail of the $\eR_{[\Ed_1, \Ed_2]}(v)$ function,  and this tail depends on the low energy tail of the resolution function $G_T(\ER, \Ed)$ in \Eq{eq:HT}, which is never well known.  Therefore too large values of $r$ make the procedure very sensitive to the way in which the tails of the  $G_T(\ER, \Ed)$ function are modeled.  This is explained in more detail in \Sec{measurements-MDM} (see also \Fig{fig:responsefunctioncogent}), where we use this procedure for a particular interaction. 
In the figures, the  horizontal placement of the vertical bar in the crosses corresponds to the maximum of $v^{-r} \, \eR_{[\Ed_1, \Ed_2]}(v)$. The extension of the vertical bar, unless otherwise indicated, shows the 1$\sigma$ interval around the central value of the measured rate.

The upper limit on the unmodulated part of $v^r \tilde{\eta}$ is simply $v^r \tilde{\eta}^{\rm lim}(v)$, where $\tilde{\eta}^{\rm lim}(v)$ is computed as described at the end of \Sec{haloindep-SI} by using a downward step-function $\tilde{\eta}_0 \, \theta(v_0 - \vmin)$ for $\tilde{\eta}^0(v_0)$ to determine the maximum value of the step $\tilde{\eta}_0$. Given the definition of the response function $\eR$ in the general case in terms of $\eH$, \Eq{eq:RT}, the downward step function choice for  $\tilde{\eta}^0$ yields
\beq
R_{[\Ed_1, \Ed_2]} = \tilde{\eta}_0 \int_0^{v_0} \ud \vmin \, \eR_{[\Ed_1, \Ed_2]}(\vmin) = \tilde{\eta}_0 \, \eH_{[\Ed_1, \Ed_2]}(v_0) ,
\eeq
From this equation we find the maximum value of $\tilde{\eta}_0$ at $v_0$ allowed by the experimental upper limit on the unmodulated rate $R^{\rm lim}_{[\Ed_1, \Ed_2]}$,
\beq
\tilde{\eta}^{\rm lim}(v_0) = \frac{R^{\rm lim}_{[\Ed_1, \Ed_2]}}{\int_0^{v_0} \ud \vmin \, \eR_{[\Ed_1, \Ed_2]}(\vmin)}
= \frac{R^{\rm lim}_{[\Ed_1, \Ed_2]}}{\eH_{[\Ed_1, \Ed_2]}(v_0)} .
\eeq

 In the figures, rather than drawing the new averages $\overline{v^r \tilde{\eta}^{\,i}}$ and the limits $v^r \tilde{\eta}^{\rm lim}(v)$, we prefer to draw $v^{-r} \overline{v^r \tilde{\eta}^{\,i}}$ and $\tilde{\eta}^{\rm lim}(v)$, so that a comparison can be easily made with the previous literature on the SI halo-independent method.

\section{Magnetic-dipole dark matter (MDM)}
\label{sec:magneticDM}

We now apply our new generalized method to a Dirac fermion DM candidate that interacts only through a magnetic dipole moment $\lambda_\chi$, the so-called magnetic-dipole dark matter (MDM)~\cite{Sigurdson:2004zp, Barger:2010gv, Chang:2010en, Cho:2010br, Heo:2009vt, Gardner:2008yn, Masso:2009mu, Banks:2010eh, Fortin:2011hv, Kumar:2011iy, Barger:2012pf, DelNobile:2012tx, Cline:2012is, Weiner:2012cb, Tulin:2012uq, Cline:2012bz},
\beq
\label{magneticDMlagrangian}
\Lag_{\rm int} = \frac{\lambda_\chi}{2} \, \bar\chi \sigma_{\mu \nu} \chi F^{\mu\nu} .
\eeq
The differential cross section for scattering of an MDM with a target nucleus is
\begin{multline}
\label{magneticDMsigma}
\frac{\ud \sigma_T}{\ud \ER} =
\alpha \lambda_\chi^2 \bigg\{ Z_T^2 \frac{m_T}{2 \mu_T^2} \left[\frac{1}{\vmin^2} - \frac{1}{v^2} \left( 1 - \frac{\mu_T^2}{\mDM^2} \right) \right] F_{{\rm SI}, T}^2(\ER(\vmin))
\\
+ \frac{\hat\lambda_T^2}{v^2} \frac{m_T}{m_p^2} \left( \frac{J_T+1}{3J_T} \right) F_{{\rm M}, T}^2(\ER(\vmin)) \bigg\} .
\end{multline}
 Here $\alpha = e^2 / 4 \pi$ is the electromagnetic fine structure constant, $m_p$ is the proton mass, $J_T$ is the spin of the target nucleus, and $\hat\lambda_T$ is the magnetic moment of the target nucleus in units of the nuclear magneton $ e / (2 m_p)= 0.16$ GeV$^{-1}$. The first term corresponds to the dipole-nuclear charge coupling, and the corresponding  charge form factor coincides  with  the usual spin-independent nuclear form factor $F_{{\rm SI}, T}(\ER)$.  We take it to be the Helm form factor~\cite{Helm:1956zz} normalized to $F_{{\rm SI}, T}(0)=1$. The second  term, which we call ``magnetic", corresponds to the coupling of the DM magnetic dipole to the magnetic field of the nucleus, and the corresponding nuclear form factor is the nuclear magnetic form factor $F_{{\rm M}, T}(\ER) $. This magnetic form factor is not identical to the spin form factor that accompanies SD interactions, in that the magnetic form factor includes the magnetic currents due to the orbital motion of the nucleons in addition to the intrinsic nucleon magnetic moments (spins).

For the light WIMPs we consider in the following,  the magnetic term is negligible for all  the target nuclei  we consider except  Na. This term is more important for lighter nuclei, such as Na and Si, but Si has a very small magnetic dipole moment. The nuclear magnetic moment of $^{23}$Na is $\hat\lambda_{\rm Na}=$ 2.218.
We took the magnetic form factor $F_{{\rm M, Na}}^2(q)$  from Fig.~31 of Ref.~\cite{Donnelly:1984rg}, which shows the ``transverse form factor" $F_{\rm T}^2(q)$ for $^{23}$Na, defined as $F_T^2(q) = q^2 F_{{\rm M}, T}^2(q) (\hat{\lambda}_T^2 / 8 \pi m_N^2) (J_T + 1) / (3 J_T)$. Here $q = \sqrt{2 m_T \ER}$, and $m_N$ is the nucleon mass. We obtain $F_{{\rm M}, T}^2(q)$ by dividing $F_{\rm T}^2(q)$ by $q^2$ and normalizing it to $F_{\rm M, Na}^2(0) = 1$. The result is fitted by the approximate functional form $F_{\rm M, Na}^2(q) = (1 - 1.15845 \, q^2 + 0.903442 \, q^4) \exp{(-2.30722 \, q^2)}$, where $q$ is in units of fm$^{-1}$.

The spin-independent part of the differential cross section has two terms, one proportional to $1/\vmin^2$ and another with a $1/v^2$ dependence. The magnetic term also has  a $1/v^2$  dependence. Notice here the difficulty that our generalized method circumvents: had we proceeded with the same usual method to compute the rate used to get to Eq.~\eqref{R1}, we would have obtained two terms  in the rate each containing a different function of $\vmin$ multiplied by detector dependent coefficients. It would have been impossible in this way to translate a rate measurement or bound into only one of the two $\vmin$ functions. 

Notice that the function $\eH_{[\Ed_1, \Ed_2]}(v)$ has in this case a $v^2$ dependence for large values of $v$, with $\eR_{[\Ed_1, \Ed_2]}(v)$ scaling as $v$.  More precisely we have

\begin{multline}
\label{eq:eH_MDM}
\eH_{[\Ed_1, \Ed_2]}(v) =
2 \sum_T \frac{C_T}{m_T} \int_0^v \ud \vmin \, \vmin
\\
\times
\left\{ Z_T^2 \left[ \frac{v^2}{\vmin^2} - \left( 1 - \frac{\mu_T^2}{\mDM^2} \right) \right] F_{{\rm SI}, T}^2(\ER(\vmin))
+
\hat\lambda_T^2 \frac{2 \mu_T^2}{m_p^2} \left( \frac{J_T+1}{3J_T} \right) F_{{\rm M}, T}^2(\ER(\vmin))
\right\}
\\
\times
\int_{\Ed_1}^{\Ed_2} \ud\Ed \, \epsilon(\Ed) G_T(\ER(\vmin), \Ed) ,
\end{multline}
where we defined $\sigma_{\rm ref} \equiv \alpha \lambda_\chi^2$. As a consequence,
\begin{multline}
\label{eq:eR_MDM}
\eR_{[\Ed_1, \Ed_2]}(v) =
2 \, v \sum_T \frac{C_T}{m_T} \int_0^\infty \ud \vmin
\\
\times \left[ \left( Z_T^2 \frac{\mu_T^2}{\mDM^2} F_{{\rm SI}, T}^2(\ER(\vmin)) + \hat\lambda_T^2 \frac{2 \mu_T^2}{m_p^2} \left( \frac{J_T+1}{3J_T} \right) F_{{\rm M}, T}^2(\ER(\vmin)) \right) \delta(v - \vmin) \right.
\\
+ \left. \frac{2}{\vmin} \, \theta(v - \vmin) F_{{\rm SI}, T}^2(\ER(\vmin)) \right]
 \int_{\Ed_1}^{\Ed_2} \ud\Ed \, \epsilon(\Ed) G_T(\ER(\vmin), \Ed) .
\end{multline}
The denominator of \Eq{averetavr} is therefore
\begin{multline}
\int \ud v \, v^{-r} \, \eR_{[\Ed_1, \Ed_2]}(v) =
2 \sum_T \frac{C_T}{m_T} \int_0^\infty \ud \vmin \, \vmin^{-r+1}
\\
\times \left[ Z_T^2 \left( \frac{\mu_T^2}{\mDM^2} + \frac{2}{r-2} \right) F_{{\rm SI}, T}^2(\ER(\vmin))
+\hat\lambda_T^2 \frac{2 \mu_T^2}{m_p^2} \left( \frac{J_T+1}{3J_T} \right) F_{{\rm M}, T}^2(\ER(\vmin)) \right]
\\
 \int_{\Ed_1}^{\Ed_2} \ud\Ed \, \epsilon(\Ed) G_T(\ER(\vmin), \Ed) ,
\end{multline}
where $r$ can be any number larger than $2$.

\section{Data comparison for MDM}\label{measurements-MDM}

The experimental data  sets we consider are the following.

{\it DAMA.} We read the modulation amplitudes from Fig.~6 of Ref.~\cite{Bernabei:2010mq}. We consider scattering off sodium only, since the iodine component is under threshold for low mass WIMPs and a reasonable local Galactic escape velocity. We show results for one single value of the Na quenching factor: $Q_{\rm Na} = 0.30$. No channeling is included, as per Refs.~\cite{Bozorgnia:2010xy, Collar:2013gu}.

{\it CoGeNT.} We use the list of events, quenching factor, efficiency, exposure times and cosmogenic background given in the 2011 CoGeNT data release \cite{CoGeNT2011release}. We separate the modulated and unmodulated parts with a chi-square fit after binning in energy and in 30-day time intervals (we fix the modulation phase to DAMA's best fit value of $152.5$ days from January 1$^{\rm st}$). We use the acceptance shown in Fig.~20 of Ref.~\cite{Aalseth:2012if}, parametrized as $C(E) = 1 - \exp(- a E)$, with $E$ in keVee and $a = 1.21$. As in \cite{Gondolo:2012rs, DelNobile:2013cta}, in the figures we plot the unmodulated component of $\tilde{\eta}$ plus an unknown flat background $b_0$.

{\it CDMS-II.} We use the germanium data (which we call CDMS-II-Ge) from the T1Z5 detector \cite{Ahmed:2010wy}, which gives the most stringent limits at low WIMP masses. We compute the upper limit on $\tilde{\eta}^0$ using the maximum gap method \cite{Yellin:2002xd} in the range $2$ keV--$20$ keV.
We also include the  CDMS-II $95\%$ upper bound of $0.045$ events/kg/day/keV on the rate modulation amplitude  for a modulation phase equal to DAMA's in the energy range $5$ keV--$11.9$ keV~\cite{Ahmed:2012vq} and use Eq.~\eqref{averetavr} or Eq.~\eqref{averetavr2} to find an upper limit on $\overline{v^r \tilde{\eta}^{\, 1}_{[\Ed_1, \Ed_2]}}$ by imposing an upper limit on $\hat{R}^{\, 1}_{[\Ed_1, \Ed_2]}$. 
In addition, we include the recent results from the silicon detector analysis in Ref.~\cite{Agnese:2013rvf}, which we denote as CDMS-II-Si. Since the energy resolution for silicon in CDMS-II has not been measured, we use the energy resolution for Ge in Eq.~(1) of Ref.~\cite{Ahmed:2009rh}, $\sigma_{\ER}(E) = \sqrt{0.293^2 + 0.056^2 E/{\rm keV}}$ keV.
With three candidate events, we calculate the maximum gap upper limit by taking $\tilde{\eta}(v)$ as a downward step function as explained at the end of \Sec{haloindep-SI}.
Assuming the events are a DM signal, we bin the recoil spectrum in $2$ keV energy intervals, 7 to 9 , 9 to 11 and 11 to 13 keV, resulting in 1 event per bin. We use the Poisson central confidence interval of $(0.173, 3.30)$ expected events for zero background at the $68\%$ confidence level to draw error bars.

\begin{figure}[t]
\centering
\includegraphics[width=0.49\textwidth]{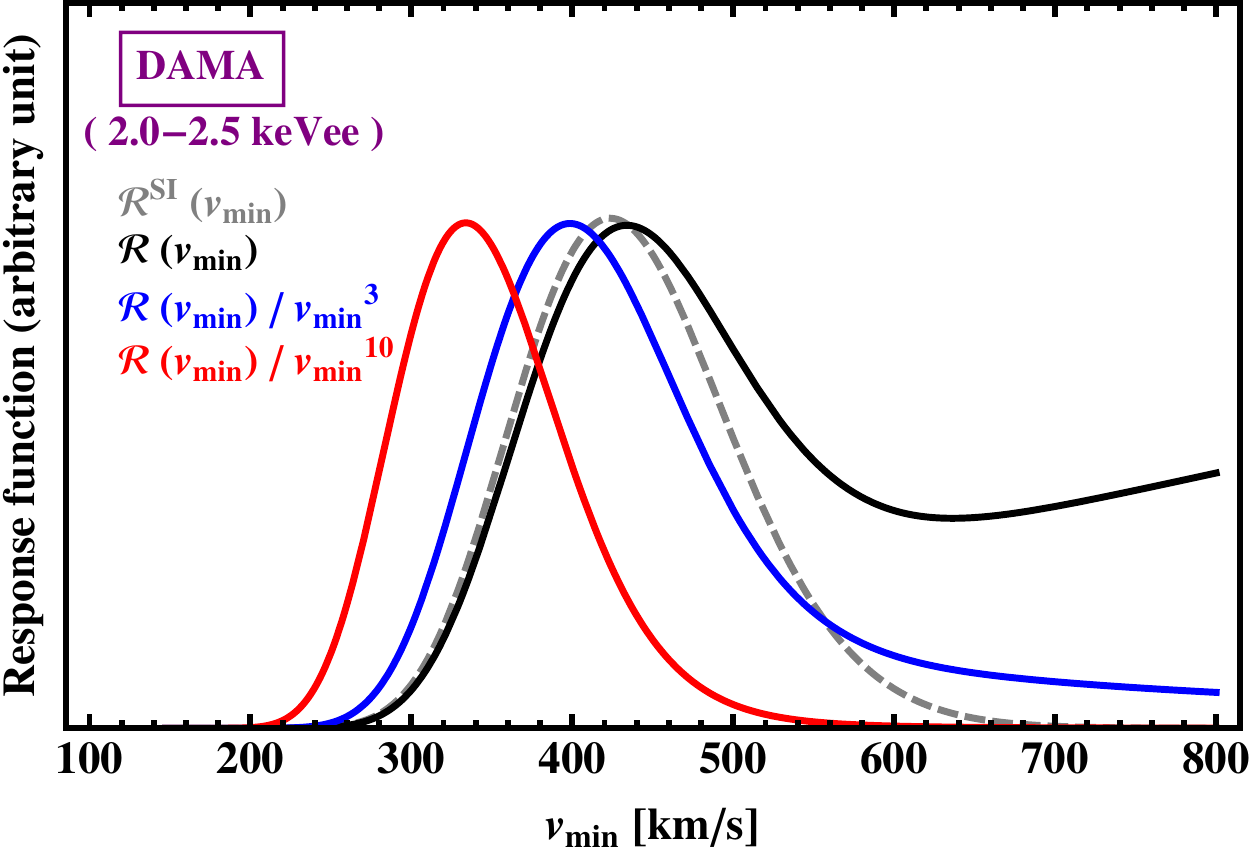}
\includegraphics[width=0.49\textwidth]{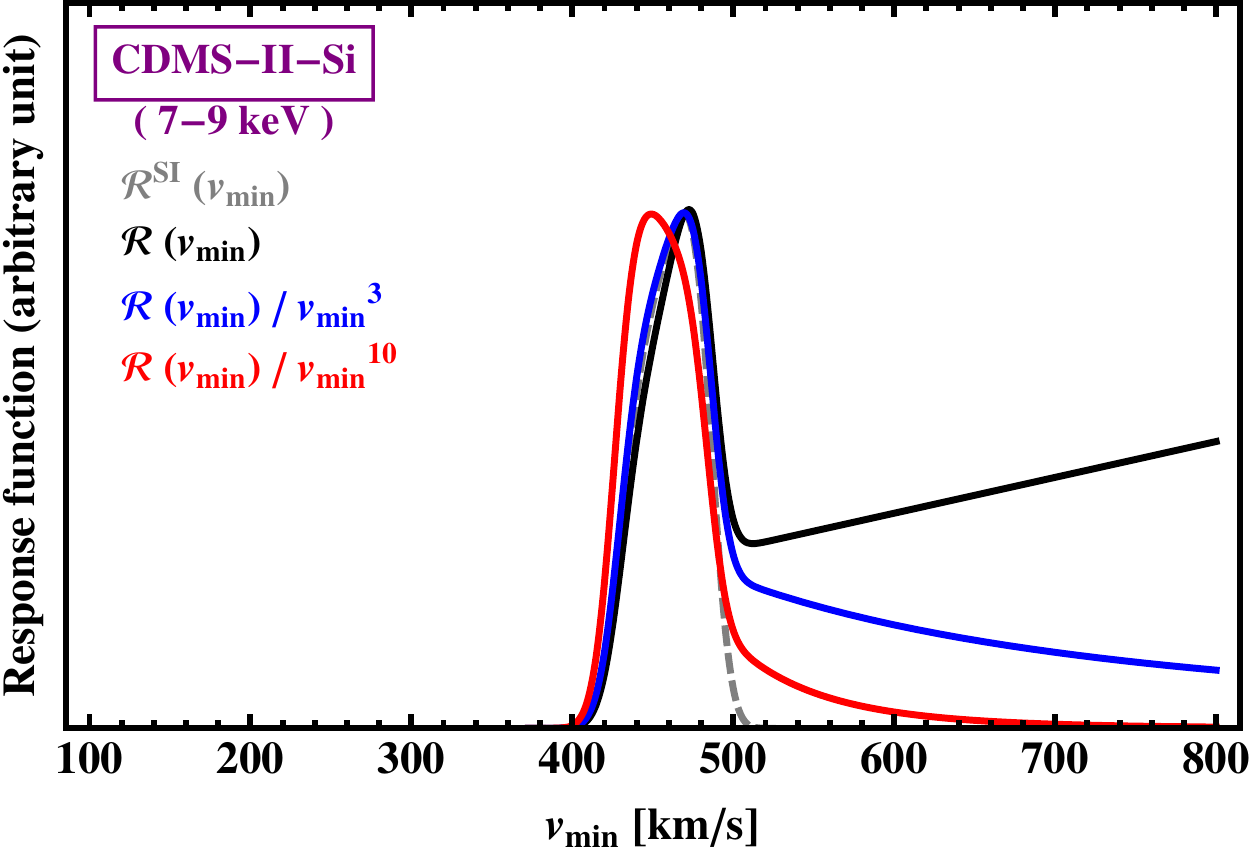}
\\
\includegraphics[width=0.49\textwidth]{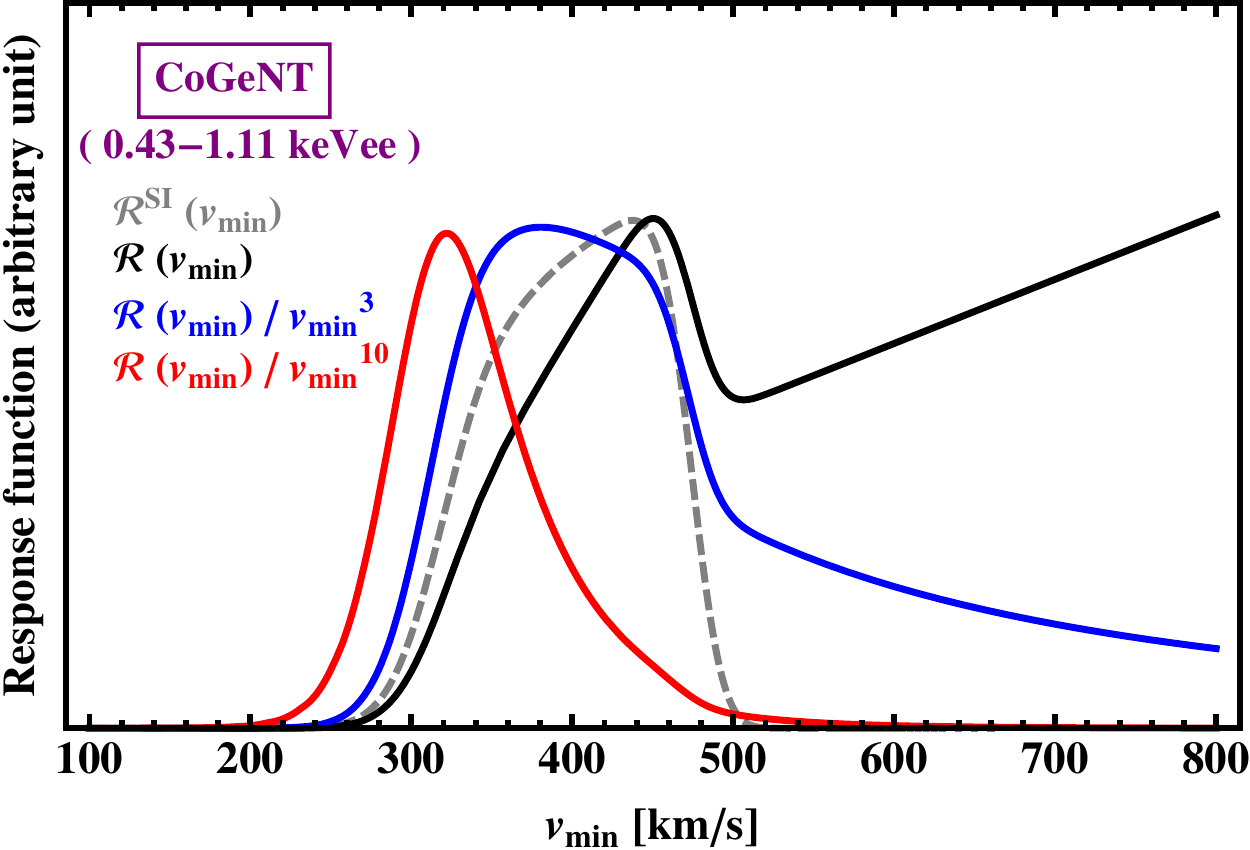}
\includegraphics[width=0.49\textwidth]{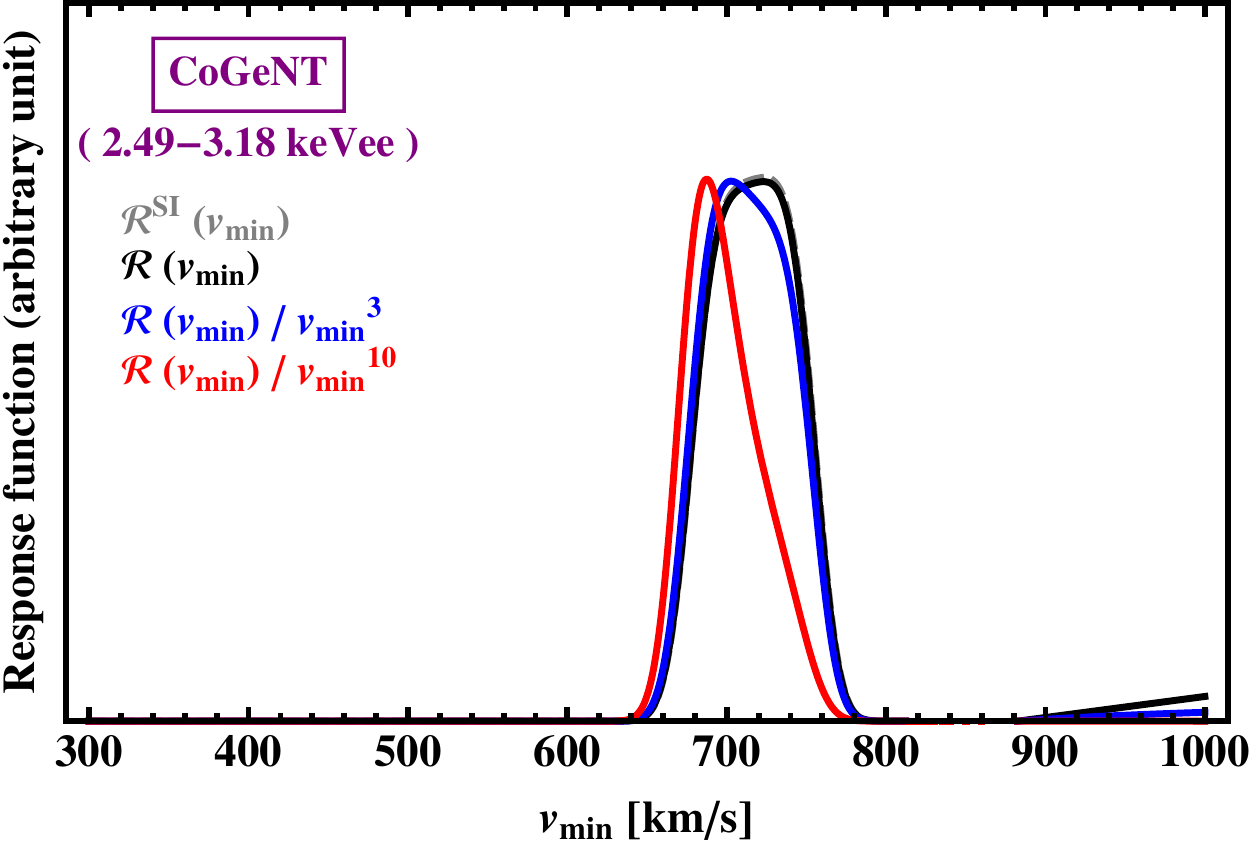}
\caption{\label{fig:responsefunctioncogent}
Response functions $\vmin^{-r} \, \eR_{[\Ed_1, \Ed_2]}(\vmin)$ with arbitrary normalization for several detected energy intervals and detectors for SI interactions (gray dashed line) and for MDM with $m = 9$ GeV.}\end{figure}

{\it XENON100.} We use the last data release of Ref.~\cite{Aprile:2012nq}, with total exposure of $224.6$ days $\times$ $34$ kg. We derive the upper limits using the expressions described in Ref.~\cite{Gondolo:2012rs}. We convert the energies of the two candidate events into $S1$ values, and use the Poisson fluctuation formula Eq.~(15) in \cite{Aprile:2011hx} to compute the energy response function. We use the light efficiency function $\mathcal{L}_{\rm eff}$ in Fig.~1 of ~\cite{Aprile:2011hi} and
the cut acceptances  of Ref.~\cite{Aprile:2012nq}.
We use the maximum gap method over the interval $3 \leqslant S1 \leqslant 30$ photoelectrons.

{\it XENON10.} We take the data from Ref.~\cite{Angle:2011th} and use only $S2$ without $S1/S2$ discrimination. The exposure is $1.2$ kg $\times$ $12.5$ days. We consider the $23$ events within the $1.4$ keV--$10$ keV acceptance box in the Phys.~Rev.~Lett.~article (not the arXiv preprint, which had an $S2$ window cut). We take a conservative acceptance of $0.94$. For the energy resolution,  we convert the quoted energies into number of electrons $n_e=E \mathcal{Q}_y(E)$, with $\mathcal{Q}_y(E)$ as in Eq.~1 of~\cite{Angle:2011th} with $k=0.11$,  and use the Poisson fluctuation formula in Eq.~(15) of \cite{Aprile:2011hx}.

\begin{figure}[t]
\centering
\includegraphics[width=0.8\textwidth]{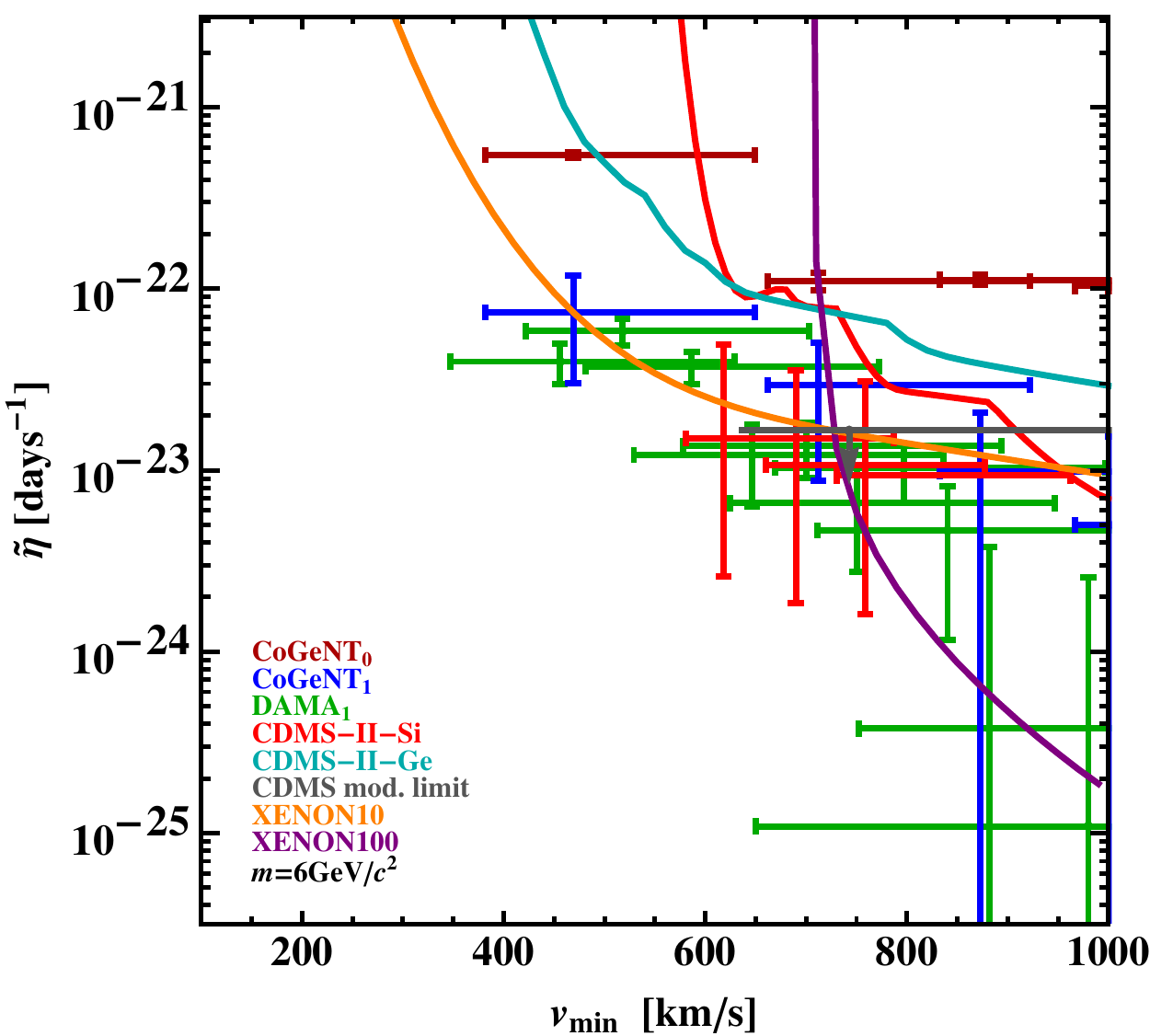}
\caption{\label{fig:2}
Measurements and bounds on $\vmin^{-10} \overline{\vmin^{10} \tilde{\eta}^{0}(\vmin)}$ and $\vmin^{-10} \overline{\vmin^{10}\tilde{\eta}^{1}(\vmin)}$ for a WIMP of mass $m = 6$ GeV with magnetic dipole interactions (MDM). The vertical axis has the usual $\tilde{\eta}$ units of day$^{-1}$.
}
\end{figure}

\begin{figure}[t]
\centering
\includegraphics[width=0.8\textwidth]{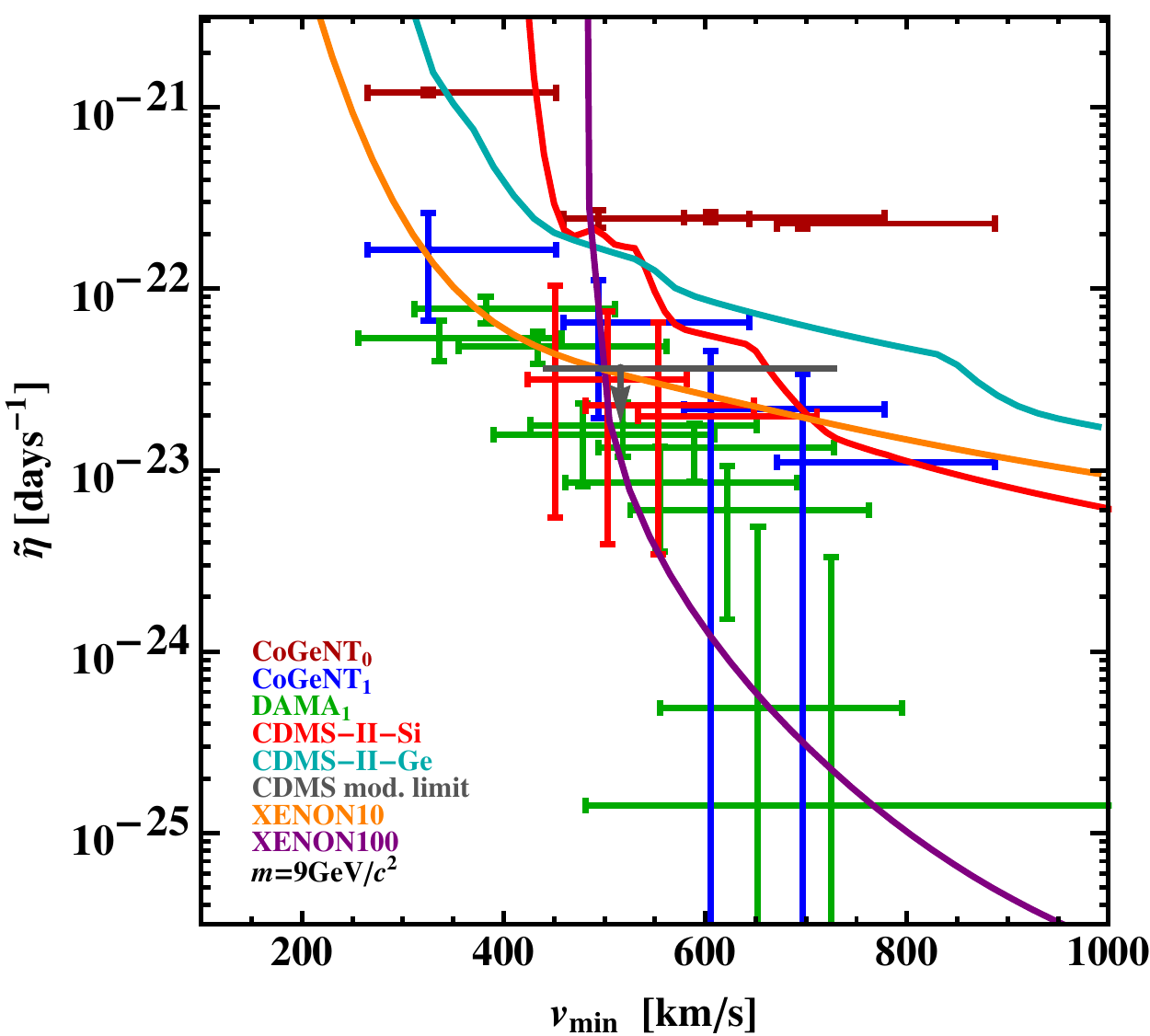}
\caption{\label{fig:3} As Fig.~\ref{fig:2} but for $m = 9$ GeV. All data points have moved to smaller $\vmin$ values as expected.
}
\end{figure}

\begin{figure}[t]
\centering
\includegraphics[width=0.8\textwidth]{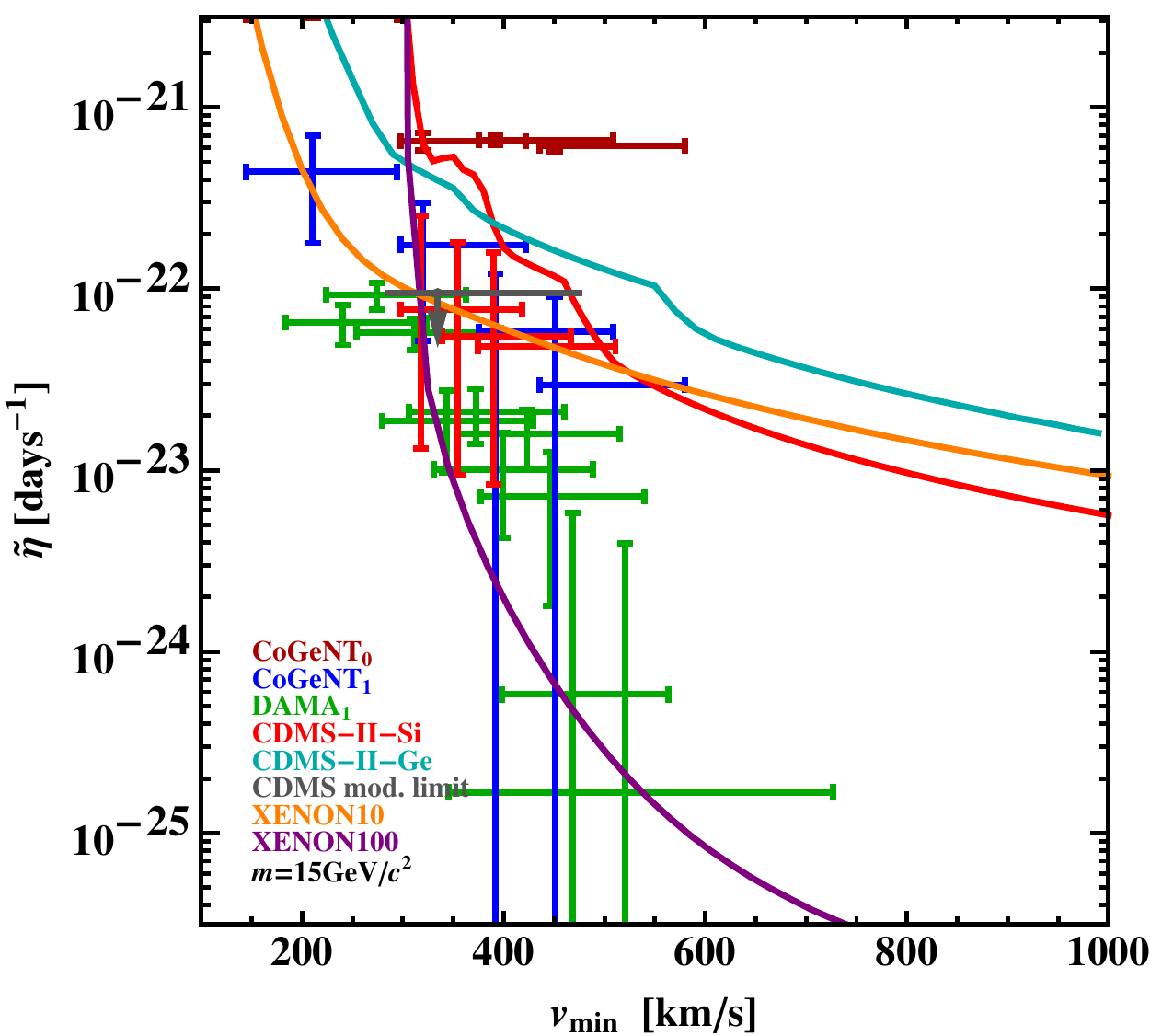}
\caption{\label{fig:4} As Fig.~\ref{fig:2} but for $m = 15$ GeV.
}
\end{figure}

In \Fig{fig:responsefunctioncogent} we illustrate the effect of various choices of $r$ on the response function $\vmin^{-r} \eR_{[\Ed_1, \Ed_2]}(\vmin)$ for MDM for several energy bins  and experiments: the first energy bin of DAMA/LIBRA~\cite{Bernabei:2010mq}, 2 to 2.5 keVee, the 7 to 9 keV CDMS-II used for the Si data~\cite{Agnese:2013rvf} and the first, 0.43 to 1.11 keVee, and last, 2.49 to 3.18 keVee, of CoGeNT~\cite{Aalseth:2010vx, Aalseth:2011wp}.
We also include $\eR^{\rm SI}_{[\Ed_1, \Ed_2]}(\vmin)$ for the standard SI interaction  (gray dashed line)  for a comparison. The normalization of each curve is arbitrary. For $r = 0$, the MDM response function is divergent and goes like $v$ at large velocities, given the $v^2$ behavior of $(v^2 \ud\sigma_T / \ud\ER)$ (see discussion at the beginning of \Sec{measurements}). The divergent behavior is much more pronounced in the low-energy bins.
The choice $r = 3$ is already enough to regularize the divergent behavior, but still yields too large $\vmin$ intervals.  For growing values of $r$,  the peak of the response function, mostly in the  low energy bins, shifts towards low velocities, due to  the $v^{-r}$ factor. This peak, when far from the  $\vmin$ interval where $\eR_{[\Ed_1, \Ed_2]}(v)$ is non-negligible, is  unreliable as it is due to the  low energy tail of the detector energy resolution function $G_T(\ER, \Ed)$, which determines the low velocity tail of $\eR_{[\Ed_1, \Ed_2]}(\vmin)$ (see \Eq{eq:HT})  and is never well known. We found the optimum $r$ value by trial an error and for MDM we find that $r = 10$  is an adequate choice (see \Fig{fig:responsefunctioncogent})  to get a localized response function in $\vmin$ space without relying on how the low energy tail of the energy resolution  function is modeled. The choice of $r$ is dictated by the lowest energy bins, where the function $v^{-r}$ is largest.
Higher energy bins are less sensitive to the choice of $r$. 

 Let us remark that  this way of comparing data is not an inherent  part to the halo independent method but only due to our choice of finding averages over measured energy bins to translate  putative measurements of a DM signal. So far we have not found a better way of presenting the data, but more work is necessary to make progress in this respect.
 
Figs.~\ref{fig:2}, \ref{fig:3} and \ref{fig:4} show the measurements and bounds on $\vmin^{-10} \overline{\vmin^{10} \tilde{\eta}^{0}(\vmin)}$ and $\vmin^{-10} \overline{\vmin^{10}\tilde{\eta}^{1}(\vmin)}$ for a WIMP with magnetic dipole interactions (MDM). To compute the position of the lines, no average is taken so that the bound corresponds to a limit on $\tilde{\eta}^0(\vmin)$.

In both figures we include the DAMA modulation signal (green crosses), CoGeNT modulated (blue crosses) and unmodulated signal (plus an unknown flat background, dark red horizontal lines), CDMS-II-Si unmodulated rate signal (red crosses and limit line), CDMS-II-Ge unmodulated rate limit (light blue line) and modulation bound (dark grey horizontal line), XENON100 225 days limit (purple line) and XENON10 S2 only limit (orange line).

Figs.~\ref{fig:2}, \ref{fig:3} and \ref{fig:4} differ for the value of the DM mass, respectively $m = 6$ GeV, $9$ GeV and $15$ GeV.  We chose these masses motivated by previous studies on MDM as a potential explanation for the putative DM signal found by DAMA/LIBRA, CoGeNT and CRESST-II (see e.g.~Ref.~\cite{DelNobile:2012tx}). The measurements and limits for MDM move to larger $\vmin$ values as the WIMP mass increases,  as expected due to the relation between $\vmin$ and the recoil energy.  As shown in Fig.~\ref{fig:2}, for a WIMP of mass $m=6$ GeV the three CDMS-II-Si points are largely below the XENON10 and XENON100 upper limits, but they move progressively above them as $m$ increases to 9 GeV, see Fig.~\ref{fig:3}, and are almost entirely excluded by them for   $m=15$ GeV in Fig.~\ref{fig:4}. The three CDMS-II-Si points overlap or are below the CoGeNT and DAMA/LIBRA measurements of the modulated part of $\tilde{\eta}$, except for the lowest energy CoGeNT and DAMA points. Thus, interpreted as a measurement of the unmodulated rate, the three CDMS-II-Si data points seem largely incompatible with the modulation of the signal observed by CoGeNT and DAMA for MDM.  For all three WIMP masses shown in the figures, the DAMA and CoGeNT modulation measurements seem compatible with each other, but the upper limits on the unmodulated part of the rate imposed by XENON10 and XENON100 reject the DAMA/LIBRA and CoGeNT modulation signal, except for the lowest energy bins, for MDM.

\section{Conclusions}

We have presented a way to generalize to any DM--target nucleus interaction the halo-independent method to compare direct dark matter results from different experiments, initially proposed in Ref~\cite{Fox:2010bz} and used already in several subsequent papers~\cite{Frandsen:2011gi, Gondolo:2012rs, Frandsen:2013cna, DelNobile:2013cta, HerreroGarcia:2011aa, HerreroGarcia:2012fu, Bozorgnia:2013hsa}. The method avoids the complications brought about by astrophysical uncertainties that affect the interaction rate.

The main idea of this method is that the interaction rate at one particular recoil energy $\ER$ depends for any experiment on one and the same  function $\eta(\vmin)$ of the minimum speed $\vmin$ required for the incoming DM particle to cause a nuclear recoil with energy $\ER$. The function $\eta(\vmin)$  depends only on the local characteristics of the dark  halo of our galaxy. Thus, all rate measurements and bounds can be translated into measurements and bounds on the unique function $\eta(\vmin)$. Before the present work, this method was applied to the standard spin-independent (SI) WIMP-nucleus interaction  only, although it could easily be applied to the standard spin-dependent (SD) interaction as well.  For both SI and SD interactions, the differential scattering cross section has a $1/v^2$ dependence on the speed $v$ of the DM particle. However, there are many other kinds of interactions with more general dependence  on the DM particle velocity and on the nuclear recoil energy and for some of them the trivial extension of the SI method does not work.  This is the case, for example when the cross section  contains two different terms with different dependences on the DM particle speed $v$.  Then, when these terms are integrated over the velocity distribution to find the rate, instead of a unique function $\eta(\vmin)$, each term has its own function of $\vmin$ multiplied by its own detector dependent coefficient. It is thus impossible  to translate a rate measurement or bound into only one of the two $\vmin$ functions contributing to the rate.

In Eq.~\eqref{R4} we have  presented a way to write the rate measured in a certain energy range  $[\Ed_1, \Ed_2]$ (expressed in observed energy $\Ed$, not in actual recoil energy $E_R$)  for any kind of interaction  in terms of a unique function of $\vmin$, which we called $\tilde{\eta}$, that depends on the local characteristics of the dark halo of our galaxy only convolved with a detector and DM candidate dependent response function in $\vmin$,  $\eR_{[\Ed_1, \Ed_2]}$.  This response function is defined in Eq.~\eqref{eq:RT}, as the derivative  of what we call the ``integrated response function"  $\eH_{[\Ed_1, \Ed_2]}$ defined in Eq.~\eqref{eq:HT} or Eq.~\eqref{eq:HT2} in terms of the scattering cross section and detector characteristics (composition, energy resolution, efficiency cuts).

Since the function $\tilde{\eta}(\vmin)$ must be common to all experiments, we can map all the rate measurements and bounds obtained with different experiments into the $(\vmin,\tilde{\eta})$ plane, as in the case of SI interactions. We have then chosen a way to compare all data for magnetic dipole moment DM (MDM) by comparing weighted averages of the $\tilde{\eta}$ derived from experiments with a potential DM signal and upper bounds on $\tilde{\eta}$ derived from data which do not find a possible DM signal. The average is weighted by the response function $\eR_{[\Ed_1, \Ed_2]}$ and corresponds to the $\vmin$ interval in which this weight function is significantly different from zero. However, we found that for a differential cross section with a general dependence on the DM velocity the width of the response function $\eR_{[\Ed_1, \Ed_2]}(v)$ in \Eq{eq:RT} at large $\vmin$,  which is dictated by the high speed behavior of the differential cross section, might even be infinite. For example, if $v^2 \, (\ud\sigma_T / \ud\ER)$ goes as $v^n$, with $n$ a positive integer,  for large $v$,
then $\eH_{[\Ed_1, \Ed_2]}(v)$ also goes as $v^n$ and  $\eR_{[\Ed_1, \Ed_2]}(v)$ goes as $v^{n-1}$ for large $v$. Thus, if $n \geqslant 1$, the response function $\eR_{[\Ed_1, \Ed_2]}(v)$ does not vanish for large $v$.  
However, we can regularize the behavior of the response function at large $v$ by using for example the function $v^r \tilde{\eta}(v)$ with integer $r \geqslant n$, instead of just $\tilde{\eta}(v)$.  Since this new function is common to all experiments, we can use it to compare the data in $\vmin$ space. The power $r$  cannot be chosen arbitrarily large, because  large values of $r$ give a large weight to the low velocity tail of the $\eR_{[\Ed_1, \Ed_2]}(v)$ function,  and this tail depends on the low energy tail of the experimental energy resolution function $G_T(\ER, \Ed)$ in \Eq{eq:HT}, which is never well known.  Therefore too large values of $r$ make the procedure very sensitive to the way in which the tails of the  $G_T(\ER, \Ed)$ function are modeled. For the particular example of interaction we present in this paper, magnetic dipole DM or MDM, we found that an optimal choice is $r=10$.
 In the figures, rather than drawing the new averages $\overline{v^r \tilde{\eta}^{\,i}}$ and the limits $v^r \tilde{\eta}^{\rm lim}(v)$, we prefer to draw $v^{-r} \overline{v^r \tilde{\eta}^{\,i}}$ and $\tilde{\eta}^{\rm lim}(v)$, so that a comparison can be easily made with the previous literature on the SI halo-independent method. 

 Let us remark that  this way of comparing data is not an inherent  part to the halo independent method but only due to our choice of finding averages over measured energy bins to translate  putative measurements of a DM signal. So far we have not found a better way of presenting the data, but more work is necessary to make progress in this respect.

\section*{Acknowledgments}
P.G.~was supported in part by NSF grant PHY-1068111. E.D.N., G.G.~and J.-H.H.~were supported in part by DOE grant DE-FG02-13ER42022.
J.-H.H.~was also partially supported by Spanish Consolider-Ingenio MultiDark (CSD2009-00064).


\appendix

\section*{Appendix A - Inelastic scattering}
The DM particle may collide inelastically with  the target nucleus~\cite{TuckerSmith:2001hy}, in which case the DM particle scatters to a different state with mass $m' = m + \delta$. Dark matter interacting inelastically via a magnetic dipole moment interaction~\cite{Chang:2010en, Kumar:2011iy} would require a modification of some of the equations presented above, in particular the definitions of  $\eH_{[\Ed_1, \Ed_2]}$. Here we present the relevant equations for the inelastic case.

In inelastic scattering, the minimum velocity the DM must have to impart a nuclear recoil energy $\ER$ depends on the mass splitting $\delta$,
\beq
\vmin = \frac{1}{\sqrt{2 m_T \ER}} \left| \frac{m_T \ER}{\mu_T} + \delta \right| ,
\eeq
where $\delta$ can be either positive (endothermic scattering~\cite{TuckerSmith:2001hy}) or negative (exothermic~\cite{Graham:2010ca}) ($\delta = 0$ for elastic scattering). Inverting this equation implies the existence of both a maximum and a minimum recoil energy for a fixed DM velocity $v$: $\ER^-(v) < \ER < \ER^+(v)$, with
\beq
\ER^\pm(v) = 
\frac{\mu_T^2 v^2}{2 m_T} \left( 1 \pm \sqrt{1 - \frac{2 \delta}{\mu_T v^2}} \right)^2 .
\eeq
The  event rate in a detected energy interval $[\Ed_1, \Ed_2]$ is  (as  in \Eq{R})
\begin{multline}
\label{R-inelastic}
R_{[\Ed_1, \Ed_2]}(t) =
\frac{\rho}{\mDM} \sum_T \frac{C_T}{m_T} \int_0^\infty \ud \ER \, \int_{v \geqslant v_\text{min}(\ER)} \hspace{-18pt} \ud^3 v \, f(\bfv, t) \, v \, \frac{\ud \sigma_T}{\ud \ER}(\ER, \bfv)
\\
\times
 \int_{\Ed_1}^{\Ed_2} \ud\Ed \, \epsilon(\Ed) G_T(\ER, \Ed)
.
\end{multline}
Changing the order of the integrations in $\bfv$ and $\ER$ in \Eq{R-inelastic}, we have
\begin{multline}\label{R6}
R_{[\Ed_1, \Ed_2]}(t) =
\frac{\rho \sigma_{\rm ref}}{\mDM} \int_{v \geqslant \hat{v}_\delta} \ud^3 v \, \frac{f(\bfv, t)}{v}
\sum_T \frac{C_T}{m_T} \int_{\ER^-(v)}^{\ER^+(v)} \ud \ER \, \frac{v^2}{\sigma_{\rm ref}} \frac{\ud \sigma_T}{\ud \ER}(\ER, \bfv)
\\
\times
 \int_{\Ed_1}^{\Ed_2} \ud\Ed \, \epsilon(\Ed) G_T(\ER, \Ed) ,
\end{multline}
where $\hat{v}_\delta$ is the minimum value $\vmin $ can take, 
$\hat{v}_\delta = \sqrt{2 \delta / \mu_T}$ for $\delta > 0$ and $\hat{v}_\delta = 0$ for $\delta \leqslant 0$. In compact form, \Eq{R6} reads
\beq
R_{[\Ed_1, \Ed_2]}(t) =  \int_{v \geqslant \hat{v}_\delta} \ud^3 v \, \frac{\tilde{f}(\bfv, t)}{v} \, \eH_{[\Ed_1, \Ed_2]}(\bfv) ,
\eeq
where as in \Eq{ftilde}
\beq
\tilde{f}(\bfv, t) \equiv \frac{\rho \sigma_{\rm ref}}{\mDM} \, f(\bfv, t)
\eeq
and
\begin{multline}
\eH_{[\Ed_1, \Ed_2]}(\bfv) \equiv
\sum_T \frac{C_T}{m_T} \int_{\ER^-(v)}^{\ER^+(v)} \ud \ER \, \frac{v^2}{\sigma_{\rm ref}} \frac{\ud \sigma_T}{\ud \ER}(\ER, \bfv)
\\
\times
 \int_{\Ed_1}^{\Ed_2} \ud\Ed \, \epsilon(\Ed) G_T(\ER, \Ed) .
\end{multline}
We will deal in  detail with the halo independent comparison of direct detection data for dark matter with magnetic dipole interactions (MDM) scattering inelastically~\cite{Chang:2010en, Kumar:2011iy} elsewhere.

\section*{Appendix B\,-\,Rate in terms of partial moments}
In this appendix we derive Eqs.~\eqref{R5} and \eqref{partialmoments}.
Define, as in \Eq{etak}, 
\begin{align}
\tilde{\eta}_{(0)}(v) & \equiv \tilde{\eta}(v) = \int_{\vmin}^\infty \ud v \, \frac{\widetilde{F}(v)}{v} ,
\label{eq:B1}
\\
\tilde{\eta}_{(k)}(v) & \equiv k  \int_v^\infty \ud v' \, \tilde{\eta}_{(k-1)}(v') , \quad \text{for integer $k>0$.}
\end{align}
Then
\begin{align}
\tilde{\eta}_{(k-1)}(v) = - \frac{1}{k} \frac{\partial \tilde{\eta}_{(k)}(v)}{\partial v} .
\end{align}
Repeatedly integrating by parts \Eq{R4} gives
\begin{align}
R_{[\Ed_1, \Ed_2]} & =  \int_0^\infty \ud v \, \tilde{\eta}(v) \, \eR_{[\Ed_1, \Ed_2]}(v) =  \int_0^\infty \ud v \, \tilde{\eta}_{(0)}(v) \, \eR_{[\Ed_1, \Ed_2]}(v) \nonumber
\\ & = -  \int_0^\infty \ud v \, \frac{\partial \tilde{\eta}_{(1)}(v)}{\partial v} \, \eR_{[\Ed_1, \Ed_2]}(v) 
=  \int_0^\infty \ud v \, \tilde{\eta}_{(1)}(v) \, \frac{\partial \eR_{[\Ed_1, \Ed_2]}(v)}{\partial v} \nonumber
\\ & = - \frac{1}{2} \int_0^\infty \ud v \, \frac{\partial \tilde{\eta}_{(2)}(v)}{\partial v} \, \frac{\partial \eR_{[\Ed_1, \Ed_2]}(v)}{\partial v}
= \frac{1}{2} \int_0^\infty \ud v \, \tilde{\eta}_{(2)}(v) \, \frac{\partial^2 \eR_{[\Ed_1, \Ed_2]}(v)}{\partial v^2} \nonumber
\\ & = \cdots \nonumber
\\ & =\frac{1}{k!}  \int_0^\infty \ud v \, \tilde{\eta}_{(k)}(v) \, \frac{\partial^{k} \eR_{[\Ed_1, \Ed_2]}(v)}{\partial v^{k}} .
\end{align}
In deriving this result, all boundary terms vanish because we have assumed that the response function and all of its derivatives vanish at $v=0$, since $v=0$ is below the threshold of any experiment. 

Notice that we can write each $\tilde{\eta}_{(k)}(\vmin)$ in terms of a single integral of ${\widetilde{F}(v)}/{v}$ as
\begin{align}
\tilde{\eta}_{(k)}(\vmin) = \int_{\vmin}^\infty \ud v \, (v-\vmin)^k \, \frac{\widetilde{F}(v)}{v} .
\end{align}
This follows by induction from Eq.~(\ref{eq:B1})  and
\begin{align}
\tilde{\eta}_{(k)}(v) &= k \int_v^\infty \ud v_2 \, \tilde{\eta}_{(k-1)}(v_2) = k \int_v^\infty \ud v_2 \int_{v_2}^\infty \ud v_1 \, (v_1-v_2)^{k-1} \, \frac{\widetilde{F}(v_1)}{v_1}
\nonumber
\\ &=
k  \int_v^\infty \ud v_1 \int_{v}^{v_1} \ud v_2 \, (v_1-v_2)^{k-1} \frac{\widetilde{F}(v_1)}{v_1} =
k  \int_v^\infty \ud v_1 \int_{0}^{v_1-v} \ud v_3 \, v_3^{k-1} \frac{\widetilde{F}(v_1)}{v_1}
\nonumber
\\ &=
\int_v^\infty \ud v_1 \left[ v_3^{k}\right]_0^{v_1-v} \frac{\widetilde{F}(v_1)}{v_1} = \int_v^\infty \ud v_1 \, (v_1-v)^k \, \frac{\widetilde{F}(v_1)}{v_1} .
\end{align}
In terms of the velocity distribution $\tilde{f}(\bfv)$, we see now that $\tilde{\eta}_{(k)}(\vmin)$ is the so-called $k$-th partial moment of the function $\tilde{f}(\bfv) / v$, as in \Eq{partialmoments} above,
\begin{align}
\tilde{\eta}_{(k)}(\vmin) = \int_{v \geqslant \vmin} (v-\vmin)^k \, \frac{\tilde{f}(\bfv)}{v} \, \ud^3 v .
\end{align}

\end{document}